\documentclass[prb,graphicx,floatfix,10pt,longbibliography, superscriptaddress, twocolumn]{revtex4-2}

\usepackage{amsmath}
\usepackage{gensymb}
\usepackage{textcomp}
\usepackage{graphicx}
\usepackage{multirow}
\usepackage{float}

\begin{document}

\title{\emph{Ab initio} study of (100) diamond surface spins}

\author{Jyh-Pin Chou}  
\affiliation{Department of Physics, National Changhua University of Education, No.1, Jin-De Road, Changhua City, Changhua, Taiwan}

\author{P\'eter Udvarhelyi}
\affiliation{Institute for Solid State Physics and Optics, Wigner Research Centre for Physics, Budapest, POB 49, H-1525, Hungary}
\affiliation{Department of Atomic Physics, Institute of Physics, Budapest University of Technology and Economics, M\H{u}egyetem rakpart 3., H-1111, Budapest, Hungary}

\author{Nathalie P.\ de Leon}
\affiliation{Department of Electrical and Computer Engineering, Princeton University, Engineering Quadrangle, Olden Street, Princeton, NJ 08544, United States}

\author{Adam Gali} \thanks{Electronic address: \texttt{gali.adam@wigner.hu}; Corresponding author}
\affiliation{Institute for Solid State Physics and Optics, Wigner Research Centre for Physics, Budapest, POB 49, H-1525, Hungary}
\affiliation{Department of Atomic Physics, Institute of Physics, Budapest University of Technology and Economics, M\H{u}egyetem rakpart 3., H-1111, Budapest, Hungary}

\date{\today}

\begin{abstract}
  Unpaired electronic spins at diamond surfaces are ubiquitous and can lead to excess magnetic noise. They have been observed in several studies to date,
  but their exact chemical nature is still unknown.
  We propose a simple model to explain the existence and chemical stability
  of surface spins associated with the $sp^3$ dangling bond on the (100) diamond surface using density functional theory. We find that the (111) facet, which is naturally generated at a step edge of (100)
  crystalline diamond surface, can sterically protect a spinful defect.
  Our study reveals a mechanism for annihilation of these surface spins upon annealing, consistent with recent experimental results.
  We also demonstrate that the Fermi-contact term in the hyperfine coupling is not negligible
  between the surface spins and the surrounding nuclear spins, and thus \textit{ab initio} simulation
  can be used to devise a sensing protocol where the surface spins act as reporter spins to sense
  nuclear spins on the surface. 
\end{abstract}

\maketitle

\section{Introduction}
The nitrogen-vacancy (NV) center in diamond is a promising technology for quantum sensing at the
nanometer scale due to its localization to atomic dimensions, remarkable chemical stability, and unique 
photophysics~\cite{maletinsky_robust_2012, mamin_nanoscale_2013, rondin_magnetometry_2014,
 muller_nuclear_2014}. In order to sense external targets, NV centers must be brought close to the surface.
 However, structural and chemical defects at the surface can give rise to paramagnetic species that cause rapid decoherence of shallow 
 NV centers~\cite{rosskopf_investigation_2014,myers_probing_2014,romach_spectroscopy_2015,favaro_de_oliveira_tailoring_2017,sangtawesin_surface_noise_2019}.
 Some of these defects give rise to unpaired electrons with $S=1/2$ and a $g$ factor of the free electron~\cite{grinolds_subnanometre_2014},
 and there are various proposals to use such defects as ``quantum reporter'' spins~\cite{sushkov_magnetic_2014}. 
 The double electron-electron resonance (DEER) signal associated with surface spins is diminished after annealing the diamond at 465 \,\celsius \ in an O$_2$ atmosphere~\cite{chu_coherent_2014,lovchinsky_nuclear_2016,sangtawesin_surface_noise_2019,dwyer_2022}. These annealing conditions have been shown to remove $sp^2$ hybridized carbon and result in a highly ordered
 oxygen terminated surface~\cite{fu_conversion_2010,sangtawesin_surface_noise_2019}. Furthermore, the DEER signal is also suppressed after annealing in vacuum at 650 \,\celsius \, at which temperature a surface dangling bond (DB) defect is also removed as evident in X-ray absorption~\cite{dwyer_2022}.
 These observations suggest that the surface spins might correspond to $sp^3$ DB species. Since $sp^3$ DB species can introduce deep acceptor states into the band gap of diamond they may be a source of charge and spin fluctuations at the surface, which may be transiently activated~\cite{chou_first-principles_2018, bluvstein_identifying_2019, yuan_charge_2020}. However, na\"ive models of dangling bonds
 at the (100) diamond surface would not be expected to be air-stable. DEER measurements~\cite{sushkov_magnetic_2014} could not identify the exact chemical environment of hydrogen atoms at the surface, and we further note that once Carr-Purcell-Meiboom-Gill protocols are applied in the detection scheme of the protons, signals associated with $^1$H and $^{13}$C nuclear spins can be overlapping~\cite{loretz_spurious_2015}, leading to further ambiguity in extracting direct information about the structure of the diamond surface. A clear understanding of the diamond surface spin will be crucial for further improving the sensitivity and resolution of diamond nanosensors.

In this paper, we propose a simple atomistic model for the near-surface DB defect on (100)
oxygenated diamond surface, and we characterize some of its properties by density
functional theory (DFT) calculations. We create a single $sp^3$ carbon DB defect on
the trench site with a terminating hydroxyl (OH) radical, see Fig.~\ref{fig:disordered}, that
can naturally occur at step edges of a polished (100) diamond surface. The proposed
annihilation mechanism of these surface spins is OH thermal desorption, which induces
minimum surface reconstruction. We estimate the temperature-dependent desorption rate
of OH by their DFT activation energy barriers,
and we find good agreement between experimental results on annealing $sp^3$ DBs \cite{dwyer_2022}
and the simulated OH desorption. Finally, we calculate the hyperfine
structure of four surface spin models, which provides direct information about the
interaction between surface spins and nuclear spins as a tool for supporting our 
model and is relevant for reporter spin protocols for sensing nuclear spins.

\section{Methods}

First-principles calculations based on DFT were performed using
the plane-wave based Vienna \emph{Ab Initio} Simulation package~\cite{kresse_efficient_1996}.
The interactions between the ions and valence electrons are treated by the projector augmented-wave~\cite{blochl_projector_1994} method. Constant volume relaxation using a cutoff energy of 370(740)~eV in the plane-wave expansion for the wave function (charge density) results in an equilibrium lattice parameter of 3.57~\AA, which is only 0.08\% larger than the experimental value of 3.567~\AA.
For the Brillouin-zone integration of a (2$\times$1) surface unit cell we use a 4$\times$8$\times$1 grid in the Monkhorst-Pack scheme~\cite{monkhorst_special_1976}.
The thickness of the vacuum layer is 10~\AA \ and the DBs on the surface are saturated by hydrogen atoms. We allow for all the atoms to fully relax until the forces are below 0.01~eV/\AA \ except for those at the bottom four layers, which are fixed at their respective positions of the (2$\times$1) reconstructed surface. 
The defect model is constructed in a 512-atom, (100)-oriented periodic slab model with 11 carbon layers and vacuum thickness of 27.5~\AA. We use this (6$\times$6) supercell to simulate a step diamond model with a surface spin. Here, we use mixed H/O/OH radicals terminating the diamond surface to minimize the surface states intrusion into the band gap~\cite{kaviani_proper_2014}.
We use the $\Gamma$-point to map the Brillouin-zone as we have tested that it is sufficient to ensure a good convergence in the total energy difference.
Our previous studies of defects in diamond~\cite{gali_theory_2009, deak_accurate_2010, deak_formation_2014} show that, unlike the Perdew-Burke-Ernzerhof (PBE) functional~\cite{perdew_generalized_1996}, the Heyd-Scuseria-Ernzerhof (HSE) hybrid functional~\cite{HSE} provides correct defect levels and defect-related electronic transitions within $\sim$0.1~eV to experiments, accurate hyperfine tensors~\cite{szasz_hyperfine_2013} and energy barriers~\cite{deak_challenges_2008}. Therefore, we apply the HSE functional in our calculations with full relaxation of the atomic positions in the surface region, using the same relaxation method as described above for the pristine unit cell. 
The acceptor and donor levels of the defect are obtained from self-consistent potential correction (SCPC) calculations as implemented in the VASP code~\cite{Silva_2021}. The SCPC method is desirable to achieve convergent electronic structure of the charged defects in diamond slab models as explained for the negatively charged nitrogen-vacancy defect in Ref.~\onlinecite{Silva_2021}.
The energy barriers of OH desorption are determined by using nudged elastic band (NEB) method~\cite{henkelman_climbing_2000}.
For the exchange-correlation functionals, we use gradient-corrected PBE functional in the NEB procedures. 

\begin{figure*}
\includegraphics[width=12cm]{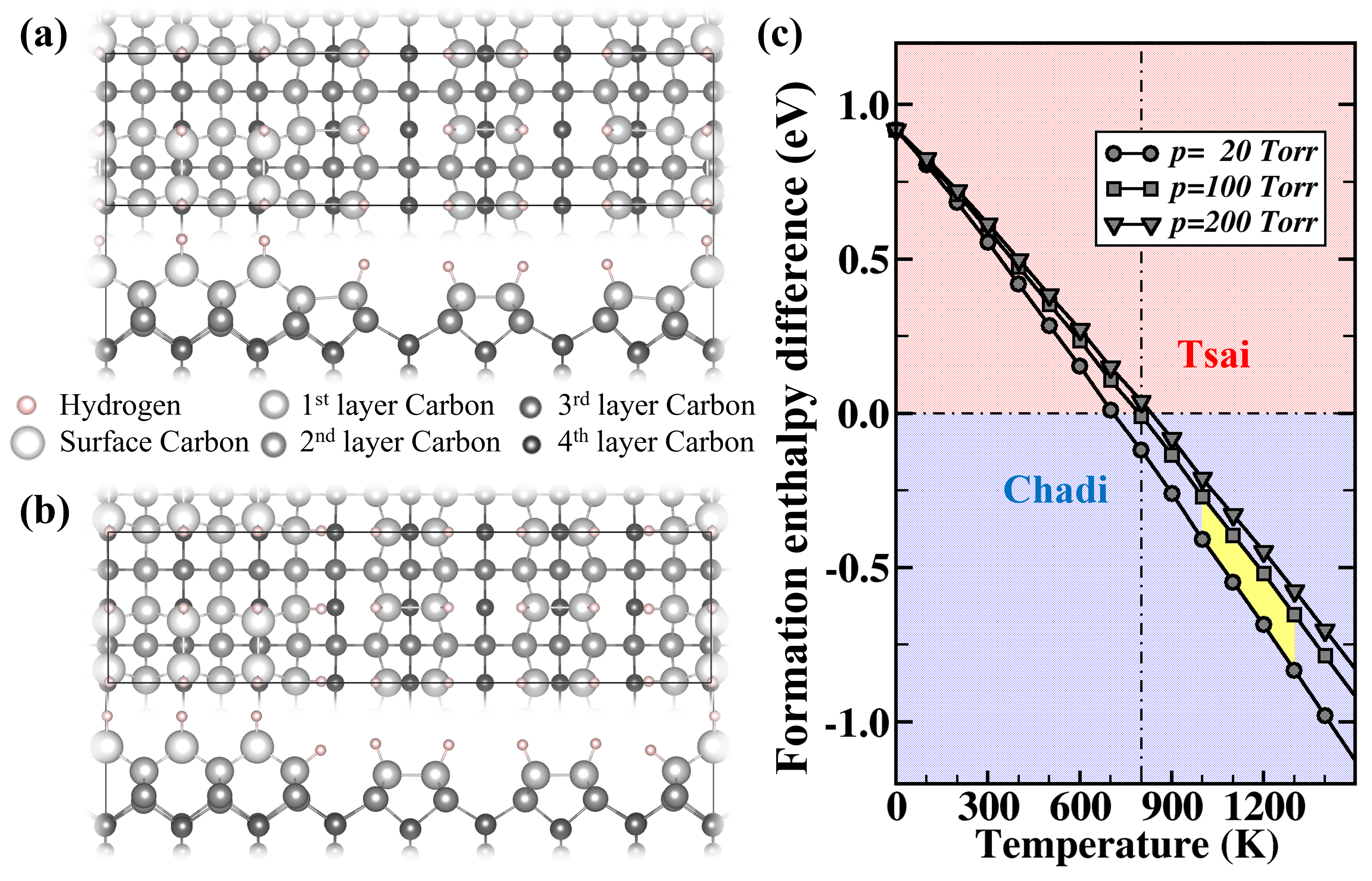}
 \caption{\label{fig:chadi_tsai} (Color online)
 Top and side views of step models of (100)-$8\times2$ diamond surface. (a) Chadi's atomic step model. (b) Tsai's atomic step model. (c) The formation enthalpy difference per unit step between Chadi's and Tsai's atomic step models as a function of hydrogen chemical potential. The red/blue region represent the Tsai's models is energetically more/less stable than Chadi's model. The typical ($p, T$) parameters are in the range of (20, 1300) $\dots$ (100, 1300) for CVD growth of diamond, marked as the yellow region.}
\end{figure*}

We also investigated the relative stability of diamond atomic step models on the (100) diamond surface on which we introduce the dangling bond defects. The two atomic structures of hydrogenated (100) diamond surface in the presence of single steps suggested by Chadi~\cite{chadi_stabilities_1987} and Tsai~\cite{Tsai_1998} are shown in Fig.~\ref{fig:chadi_tsai}(a) and (b), respectively. For the sake of clarity, the highest five layer of carbon atoms are presented in different sizes and colored in different shades of grey (higher atoms are larger and lighter grey). Essentially, in the same size of unit cell, the difference between the Chadi and the Tsai models is the number of C-C surface dimers. In a C(100)-$8\times2$ supercell, the Tsai model possesses two more C-C dimers than Chadi model, so the corresponding number of hydrogen atoms on the top of surface in the unit cell of Chadi and the Tsai models are 14 and 18. The distance between two step edges are 8.8~{\AA} in Chadi model and 12.7~{\AA} in Tsai model. To examine if the interaction between two step edges affects the formation entropy, we calculated the Chadi model with a shorter distance at 7.7~{\AA}, and the results are consistent with the longer distance. Temperature dependence of the step formation enthalpy is mainly incorporated through the hydrogen chemical potential ${\mu}_{\text{H}}$. The step formation enthalpy per step unit length of H-terminated diamond surface can be calculated as follows,
\begin{equation}
    H(p,T) = \left[E_{\text{total}} + F(T) - n_{\text{C}}{\mu}_{\text{C}} - n_{\text{H}}{\mu}_{\text{H}}(p,T) \right]/4 \text{,}
\end{equation}
where $E_{\text{total}}$ is the total energy, $F(T)$ is the free energy arising from the vibrational modes of the hydrogen atoms at the surface, ${\mu}_{\text{C}}$ is the chemical potential of carbon atom, and $n_{\text{C}}$($n_{\text{H}}$) are the numbers of carbon(hydrogen) atom. The factor of 4 arises from the fact that there are two step edges and two step unit lengths per unit cell in our models. Since the system is assumed to be in thermal equilibrium, we use the total energy of bulk diamond as the carbon chemical potential ${\mu}_{\text{C}}$. To determine $F(T)$ we calculated the local vibrational modes (LVMs) of C-H for these two models. One can calculate the $F(T)$ as
\begin{equation}
    F(T) = \sum_i \left\{\frac{E_i}{2} + k_{\text{B}}T \ln\left[ 1 - \exp\left(\frac{-E_i}{k_{\text{B}}T}\right) \right]\right\}\text{,}
\end{equation}
where $E_i=h\nu_i$ and $\nu_i$ is the frequency of the $i$th LVM, $h$ is Plank constant. $F(T=0~\mathrm{K})$ gives the zero point energy of the diamond surface. It is possible to express the chemical potential of hydrogen ${\mu}_{\text{H}}$, assuming equilibrium with the gas phase as
\begin{equation}
    {\mu}_{H}(p,T) = {\mu}_{H}(p,0~\mathrm{K}) + \frac{\Delta G(p_0,T)}{2} + \frac{k_{\text{B}}T \ln(p/p_0)}{2}\text{,}
\end{equation}
where ${\mu}_{H}(p,0~\mathrm{K})$ is the total energy of hydrogen at 0~K and $p_0$ is the standard state's pressure. In the conventional chemical vapor deposition (CVD) diamond growth, mixture of hydrogen and methane are commonly used as the reactant gas. Typically, the concentration of methane is 0.1~1.0\%. Therefore, it is reasonable to take ${\mu}_{H}(p,0~\mathrm{K})$ to be the total energy of hydrogen in an isolated molecule form, i.e.,
\begin{equation}
    {\mu}_{H}(p,0~\mathrm{K})=\frac{E_{\text{CH}_{4}} + E_{\text{CH}_{4}\text{,ZPE}} - {\mu}_{\text{C}}}{4}\text{,}
\end{equation}
where $E_{\text{CH}_{4}\text{,ZPE}}$ is the zero-point energy of $\text{CH}_4$. The second term $\Delta G(p_0,T)$ is the difference in the Gibbs free energies, which can be obtained from the differences in the enthalpy and entropy of a methane molecule with respect to the $T=0~\mathrm{K}$ limit, e.g., tabulated in the thermodynamic tables.


\section{Results}

We summarize the following facts about surface spins found in experimental results~\cite{sushkov_magnetic_2014, chu_coherent_2014, lovchinsky_nuclear_2016, bluvstein_identifying_2019, sangtawesin_surface_noise_2019, yuan_charge_2020, dwyer_2022}: 
 (i) one type of the surface spins correlates with X-ray signatures of an $sp^3$ DB,
 (ii) it is a spin-1/2 electronic defect, 
 (iii) it must be located within a few layers of the surface, and 
 (iv) it is chemically stable in air. 
 (v) Furthermore, the defect should stay neutral to remain paramagnetic during the optical measurements. This can be achieved when the occupied defect level falls relatively close to the valence band maximum whereas the unoccupied levels falls high in the gap to produce hyperdeep donor and acceptor levels.
 
The surface $sp^3$ DBs are not stable in air, thus we study a vacancy-like structure beneath the surface. Therefore, we introduce an atomic step on (100) diamond surface, which was also motivated by the experimental conditions~\cite{luo_polishing_2021, sangtawesin_surface_noise_2019} as will be explained below.

\subsection{Surface models}

In experiment, the surface layer is generally damaged and highly strained because of mechanical polishing~\cite{luo_polishing_2021}, and this strain can be mitigated by etching and surface processing~\cite{sangtawesin_surface_noise_2019}. Systematic simulation and analysis of such disordered diamond surfaces is computationally prohibitive at \emph{ab initio} level needed for 
accurate electronic structure calculation. Instead, we attempted to set up a simple
but still relevant model to identify the surface spins using the following steps.

 (i) We start with an atomically flat (100) diamond surface. It is well-known that the dangling bonds at the surface are highly reactive, thus the surface spins should be located beneath the topmost surface.

 (ii) For all carbon atoms of a pristine (100) diamond surface, there are two bonds pointed up, out of the surface and two bonds pointing downward. Removing/adding one carbon atom always generates two DBs. It is therefore not possible to construct a single $sp^3$ DB with a small number of atoms at the (100) surface.
However, at the diamond (111) surface, each carbon atom possesses three bonds downward(upward) and one bond upward(downward), thus it is possible to form a single $sp^3$ DB defect within minimum change of the diamond lattice. Therefore, to form a $sp^3$ DB beneath (100) surface, a (111) facet is essential.

 (iii) Reflection anisotropy spectroscopy ~\cite{schwitters_contribution_2011} shows that a single layer step on high-quality atomically smooth H/C(100)-2$\times$1 surfaces is predominantly realized as the Chadi step model in CVD grown diamonds~\cite{chadi_stabilities_1987}. One can find a tilted (111) facet at the (100) diamond surface caused by a single atomic step layer in the Chadi step model [see Fig.~\ref{fig:chadi_tsai}(a)]. 

Despite the experimental evidence of the Chadi step model for the H/C(100)-2$\times$1 diamond surface, an alternative Tsai step model~\cite{Tsai_1998} exists [c.f. Figs.~\ref{fig:chadi_tsai}(a) and (b)] which was predicted to be more stable than the Chadi step model by DFT calculations. In order to understand this issue, we study the relative stability of the two models with taking into account the conditions of formation. We realized that Chadi step model contains less number of C-H bonds than Tsai step model does. Since C-H bonds possess strong local vibration modes the temperature and pressure dependent formation enthalpy may significantly change the relative stability between these two structures with respect to the formation energies neglecting the zero-point-energy contributions.

The formation entropy difference per unit step length between these two models as a function of ${\mu}_{\text{H}}$ are plotted in Fig.~\ref{fig:chadi_tsai}(c). Indeed, Tsai model is 1.2~eV more stable than Chadi model at temperature is equal to 0~K which is in agreement with Tsai’s results~\cite{Tsai_1998}. However, as temperature increases the formation entropy difference becomes small and the order of relative stability shows a transition from Tsai's model to Chadi model at temperature of 800~K (at $p=100$~Torr). Microwave plasma-assisted chemical vapor deposition (MPCVD) is the major growth processes of high-quality homoepitaxial growth of diamond surface. Although CVD is a non-equilibrium process, the study of equilibrium system would be the first step in understanding the stability of different phases during the growth. In the CVD homoepitaxial growth of diamond, the surface is mostly terminated by hydrogen because abundant hydrogen is used in the growth environment. Typical growth conditions for conventional CVD diamond are under gas pressures of $20\dots 100$~Torr and substrate temperature of $1000\dots 1300$~K; for high-power MPCVD, the pressure is $100\dots 200$~Torr and substrate temperature of $1200\dots 1500$~K. Our calculation results reveal that Chadi model is more favored under CVD growth conditions. As the Chadi's single step model is energetically more stable than Tsai model at typical CVD growth conditions, $1000\dots 1300$~K and $20\dots 100$~Torr as the yellow region in Fig.~\ref{fig:chadi_tsai}(c), we can deduce that the step structure at (100) CVD diamond surface will be dominated by the formation of Chadi step model. 

We note that the above mentioned simulations are directly relevant for the hydrogenated (100) CVD diamond layers. On the other hand, the diamond surface is polished and etched after introducing the NV quantum sensor to diamond in experiments. We note that the atomistic simulations of these complex processes are out of reach with the present computational power. We assume here that local ``disorder'' with realizing (111) facet on (100) diamond surface can be modeled the most consistent way by a single atomic Chadi step model which is proven to be a stable defect species on (100) diamond surface. This model increases the complexity of the surface model at minimum level with respect to the atomically smooth surface model and can produce a topographically protected $sp^3$ DB as will be shown below.

\begin{figure*}
\includegraphics[width=16cm]{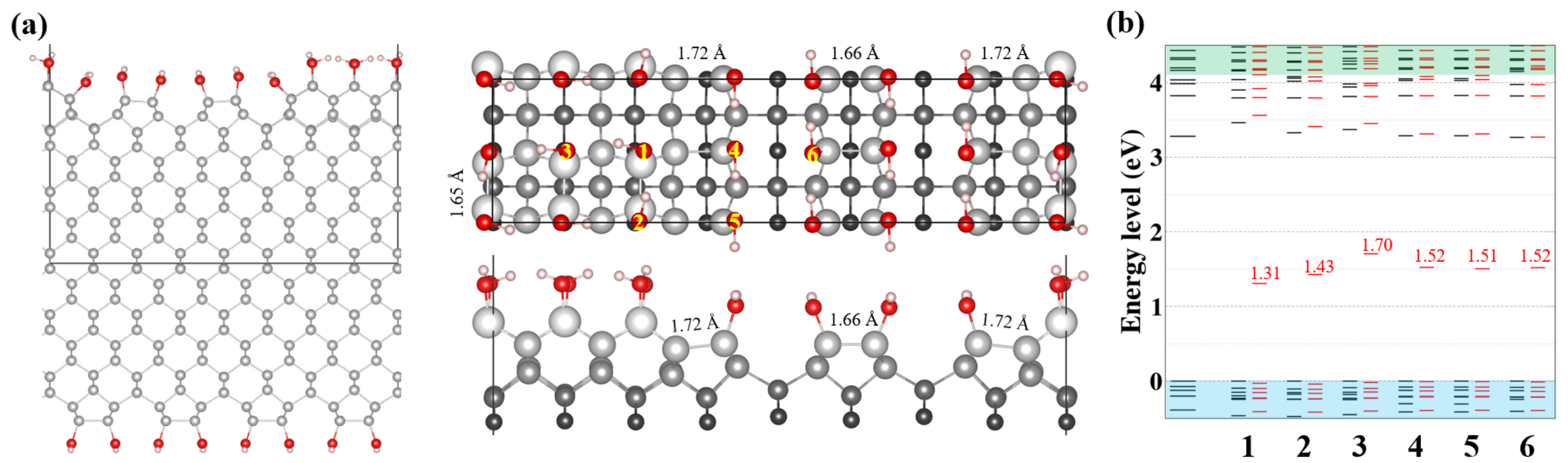}
 \caption{\label{fig:chadi_db} (Color online)
 (a) Structure of Chadi step model. Side view (left) and top view (right) are presented. For the sake of clarity, only the topmost five layers is displayed in the top view and the color scheme is consistent with Fig.~\ref{fig:chadi_tsai}(a). (b) Energy levels of surface dangling bond state. Black and red dashes represent spin majority and minority channels, respectively. Blue and green regions represent valence and conduction bands, respectively.}
\end{figure*}

\begin{figure}
\includegraphics[width=8cm]{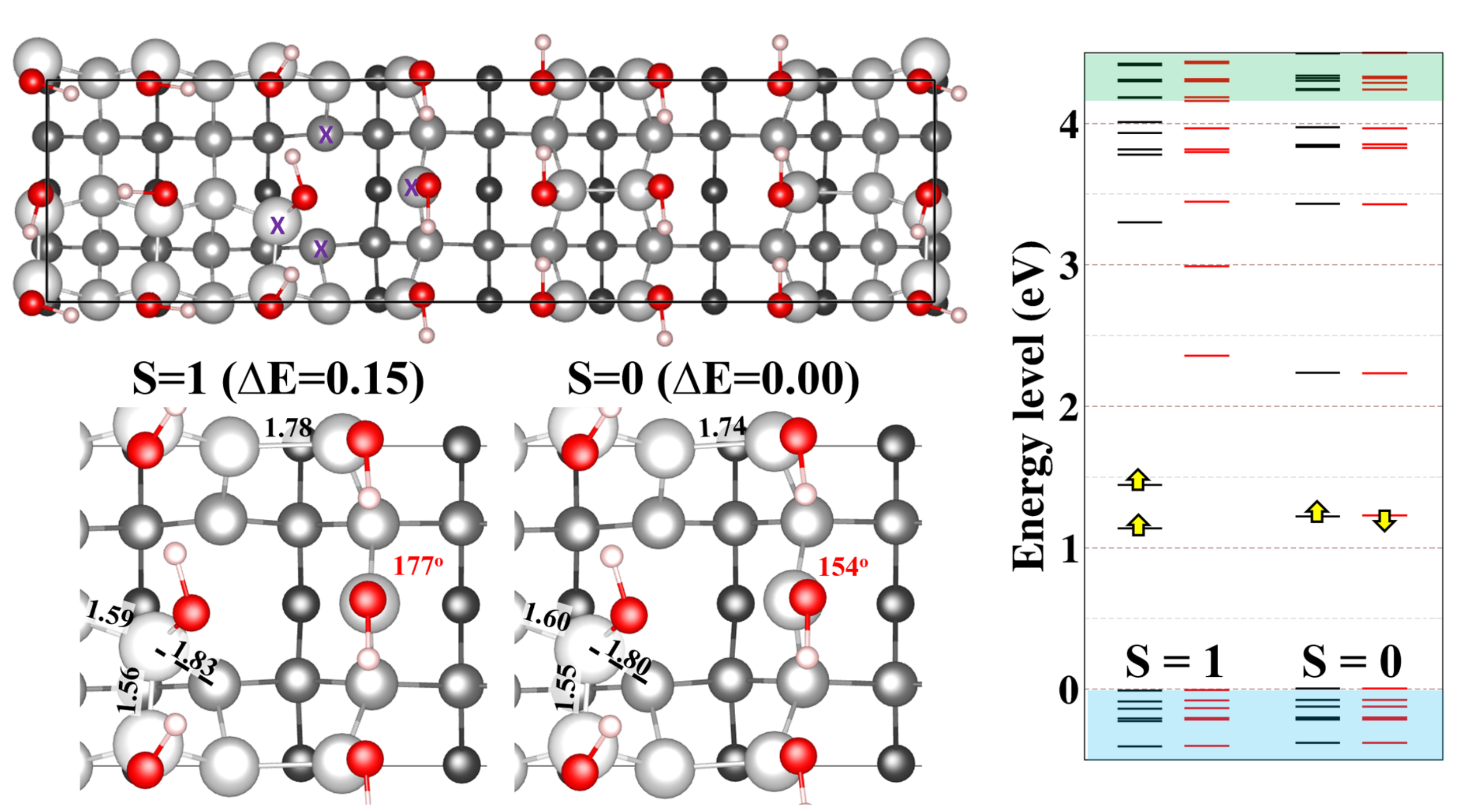}
 \caption{\label{fig:carbon_vacancy} (Color online)
 Structure of carbon vacancy at step edge and its corresponding energy level plot. The “X” denotes the carbon atoms with coordination number of 3.}
\end{figure}

\subsection{Dangling bond models}

We examine around 80 plausible prototype models that contain an $sp^3$ DB near the (100) diamond surface. We apply PBE calculations to screen the prototype models based on the five criteria listed above. The most plausible model is selected in this procedure of which stability and other properties are studied in detail with advanced DFT calculations.

\subsubsection{Surface dangling bonds on Chadi C(100) step model}

First, we investigate the dangling bond on the topmost layer. We build a OH-terminated Chadi-type diamond (100)-$8\times2$ step model with 11 carbon bilayers as shown in Fig.~\ref{fig:chadi_db}(a).
The structure (side view) is shown at left. For the sake of clarity, we present topmost few layers at right (top view) and change the color deep to show carbon atoms in different layers. Then we investigate the surface dangling bond models by removing OH radical at six different sites. The structure (right) and the corresponding electronic structure (left) are shown in Fig~\ref{fig:chadi_db}(b). In the electronic structure plot, the leftmost plot shows the energy levels of the model without surface dangling bonds. All the results indicate that the occupied dangling bond levels are located near 1.5~eV above VBM which lies to high to create a hyperdeep donor level. 

Next, we remove one carbon on the step edge. One missing carbon atom at the edge will generate two $sp^3$ dangling bonds, one $sp^2$ carbon atom, and one $sp^2$-like carbon atom, see Fig.~\ref{fig:carbon_vacancy}. The “X” denotes the carbon atoms with coordination number of 3. In this configuration, the spin configurations of $S = 0$ is more stable than $S = 1$. The optimized structures are presented in detail. The $sp^2$-like carbon can protect one $sp^3$ dangling bond. However, the other $sp^3$ dangling bond is easily attacked by the species in the air. Thus, this model is not accepted for further consideration.

\subsubsection{Single vacancy}

\begin{figure*}
\includegraphics[width=12cm]{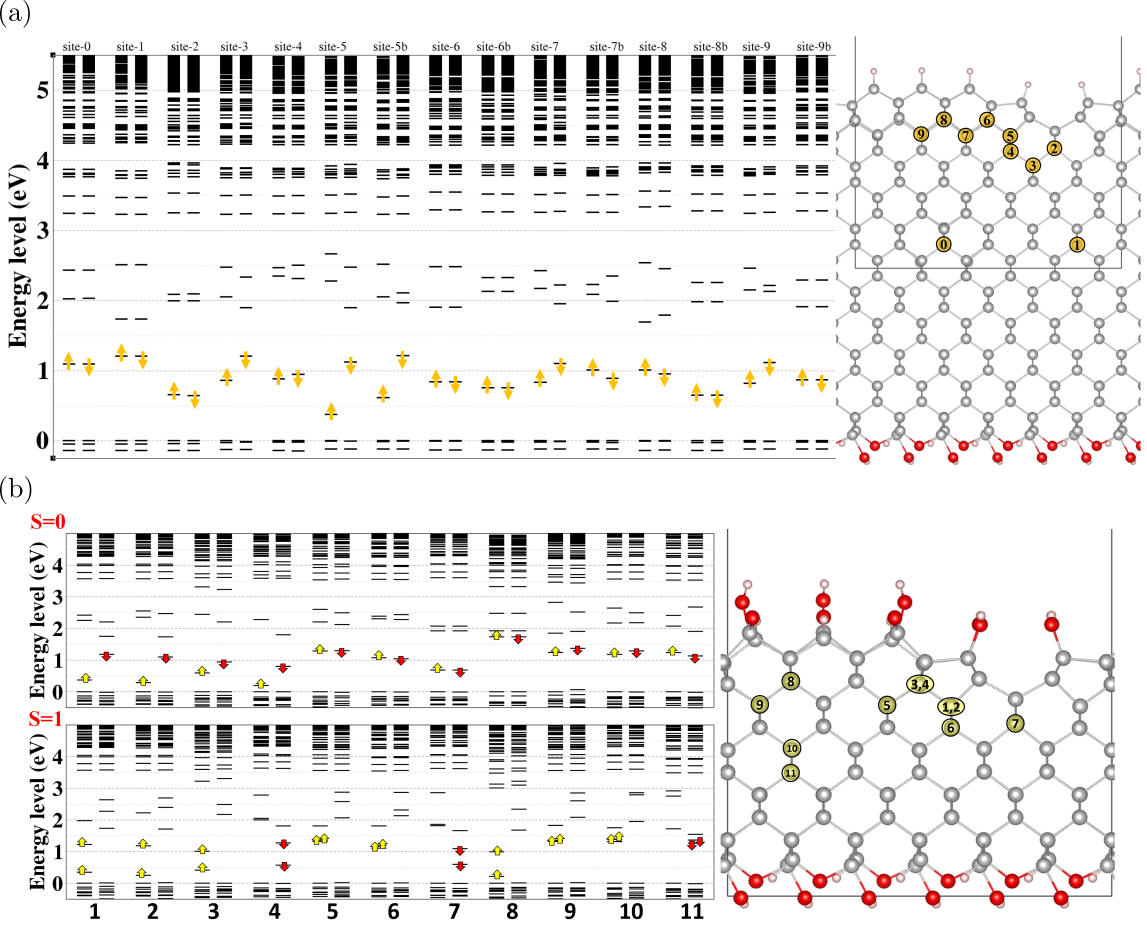}
 \caption{\label{fig:single_vacancy} (Color online)
 (a) Energy level plot for a single vacancy on the H-terminated diamond (100) surface. To compare the energy level variations of single vacancies near the surface, a diamond model with a thickness of twenty-two layers is utilized. Site-0 and site-1 are employed to simulate bulk states. (b) Energy level plot for a single vacancy on the OH-terminated diamond (100) surface. In this case, a ten-layer thickness is introduced to investigate the surface states.}
\end{figure*}

The vacancy in the diamond bulk can diffuse to the surface via heat treatment so the surface spin might originate from a single vacancy defect near the surface. To model this scenario, we use a hydrogen terminated step model and create a vacancy near the surface. The possible positions and corresponding energy levels ($S = 0$) are presented in Fig.~\ref{fig:single_vacancy}(a).
For site-5 to site-9 rows, there are two different positions in the same row which are labeled as site-5b to site-9b. We calculate both $S = 0$ and $S = 1$ states, and we found that the site-2 with spin configuration of $S = 0$ is the most stable site (at least 0.3~eV energy lower than the others) among these models. However, the defect levels fall too high in the gap. We also calculated the total energy and energy levels for the vacancy defect in this H/O/OH terminated surface. The results are shown in Fig.~\ref{fig:single_vacancy}(b). A vacancy on site-8 is the most stable one among these models. However, the site-8 configuration is beneath the surface terrace. Creating a $S = 1/2$ state on the terrace with this defect without significantly modifying the terrace structure is impossible. Therefore, a single vacancy configuration is excluded from further considerations.

\subsubsection{Single vacancy with hydrogen atoms}

\begin{figure}
\includegraphics[width=8.2cm]{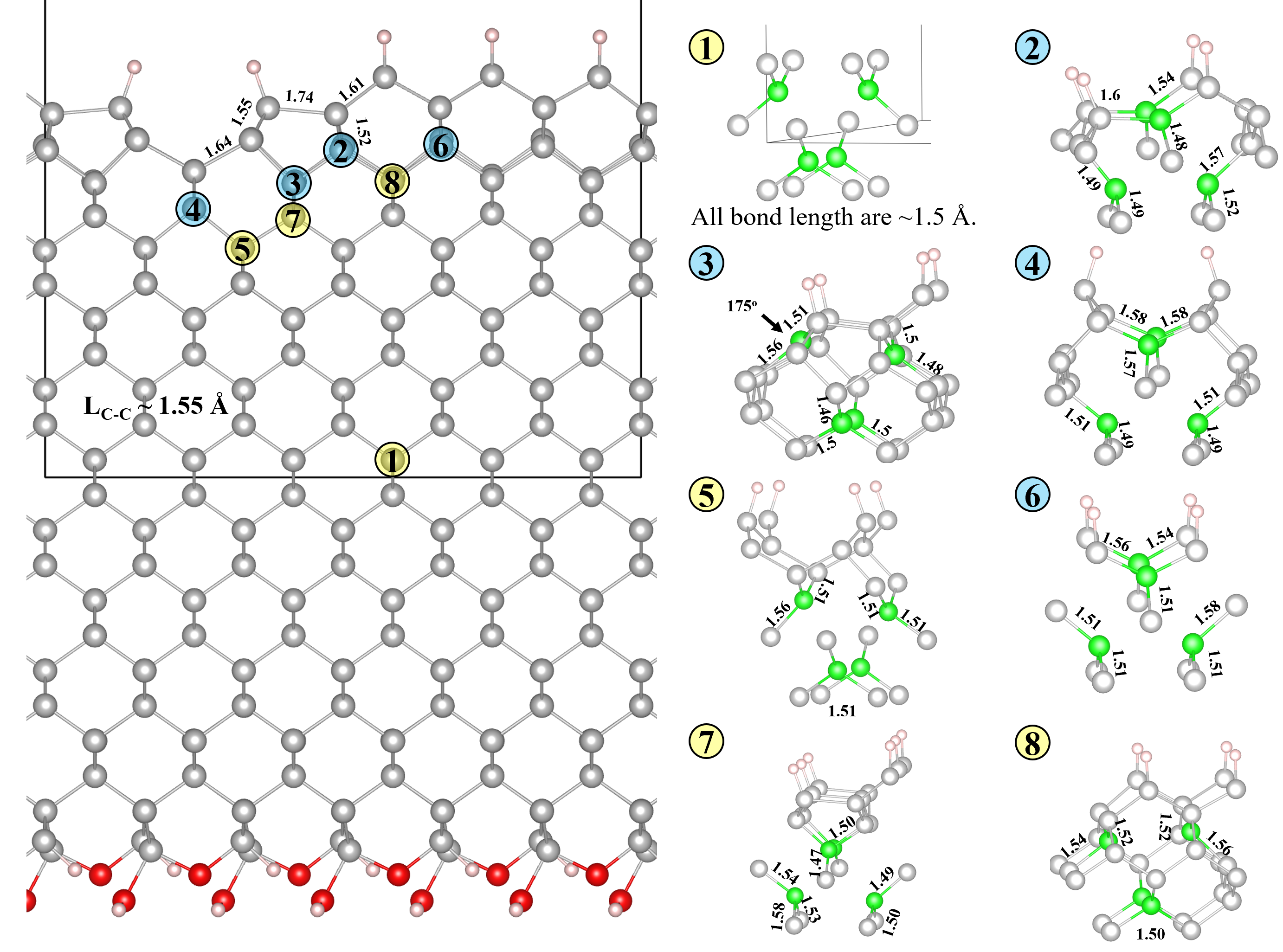}
 \caption{\label{fig:Hterm_vacancy_struct} (Color online)
 Structure details of single vacancy in the H terminated step model. The four carbon atoms surround the vacancy site are highlighted with light-green color.}
\end{figure}

\begin{figure}
\includegraphics[width=8.2cm]{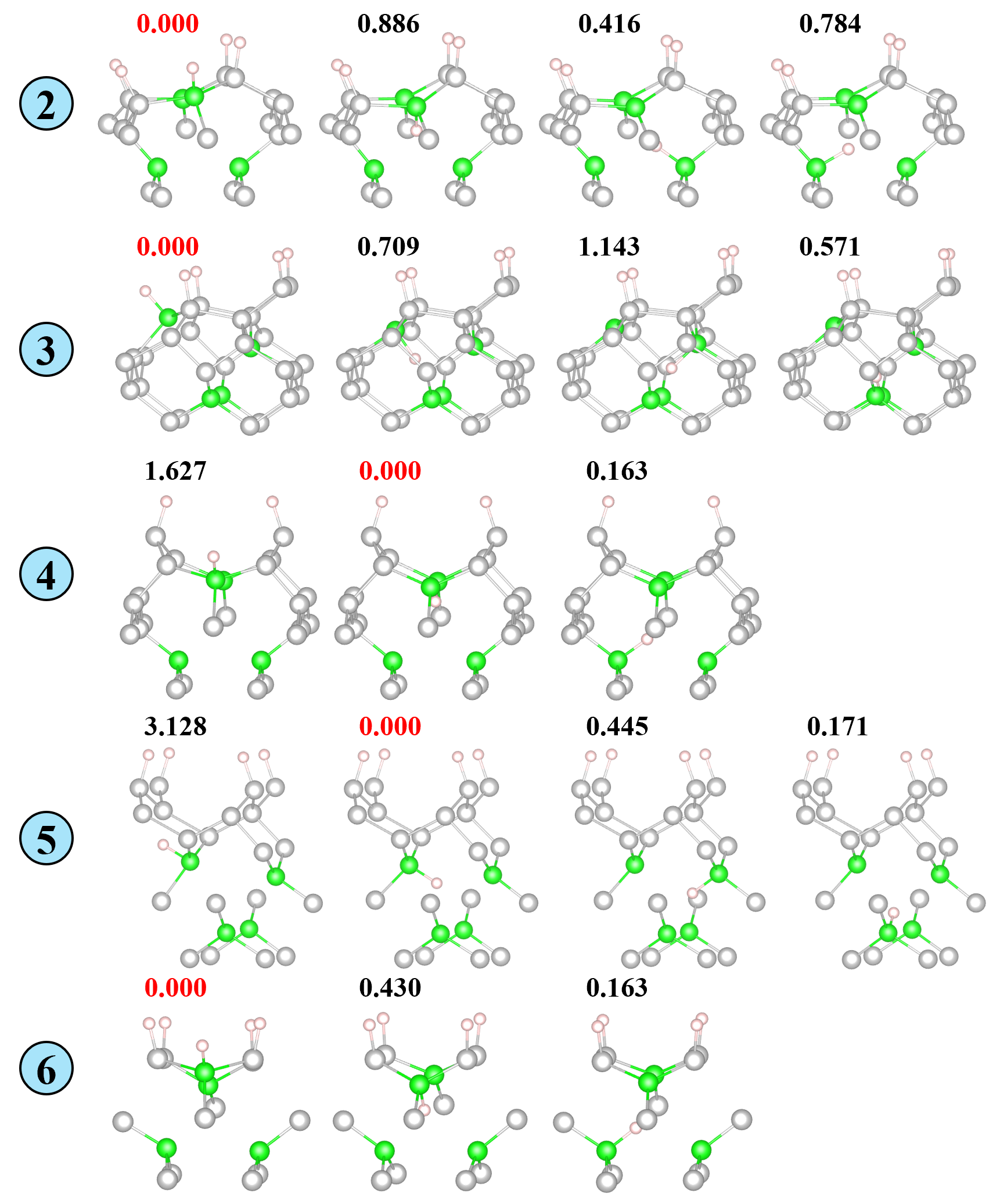}
 \caption{\label{fig:vacancy_energy} (Color online)
 The relative energy (unit in eV) of hydrogen saturation of surface vacancy. The four carbon atoms surround the vacancy site are highlighted with light-green color.}
\end{figure}

\begin{figure}
\includegraphics[width=7cm]{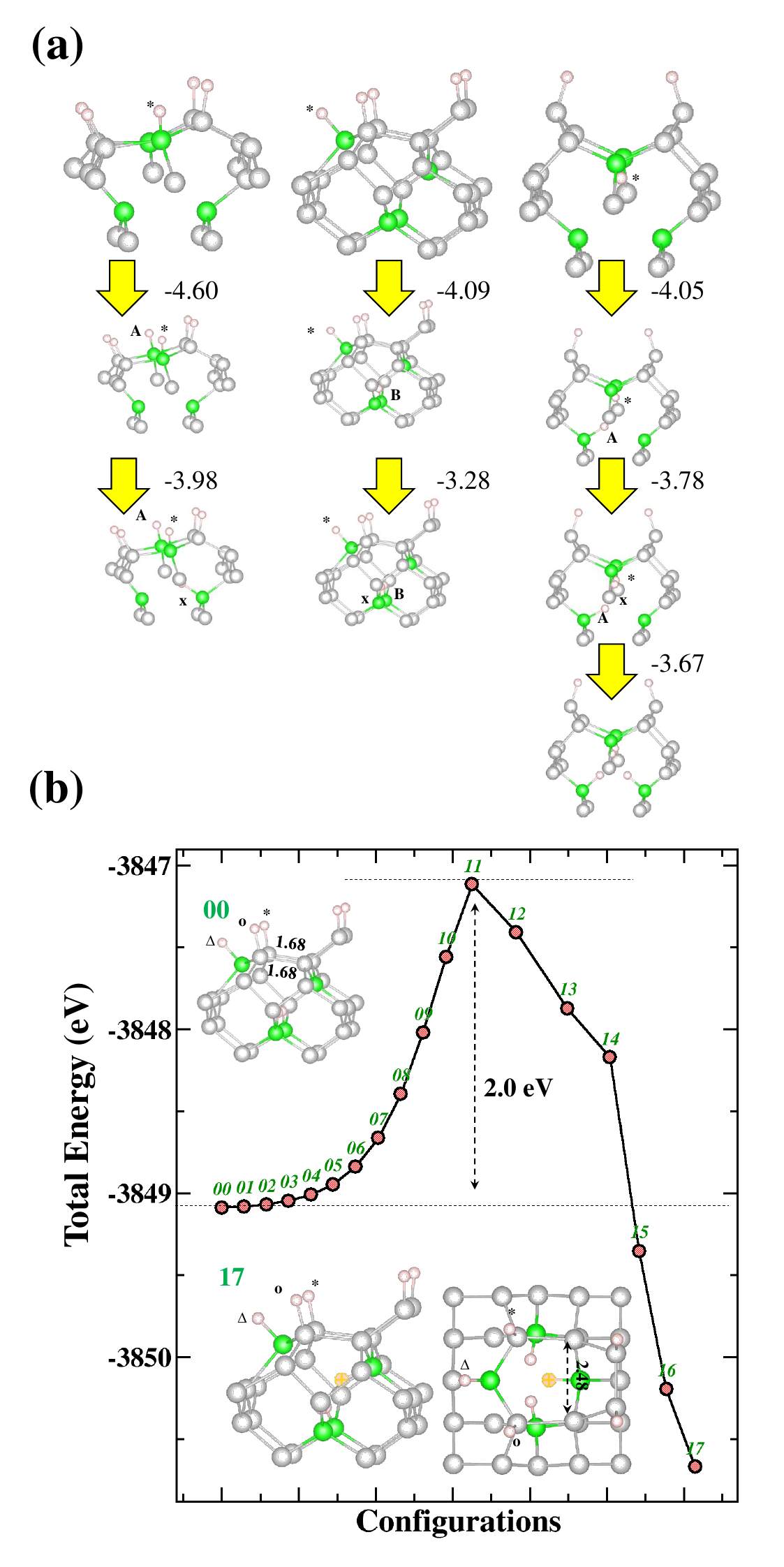}
 \caption{\label{fig:vacancy_saturation} (Color online)
 (a) Hydrogen saturation of surface vacancy. To assist with visual guidance, the additional hydrogen atoms are marked with symbols "*", "x", "A", or "B". (b) The reaction of hydrogen desorption from surface vacancy.}
\end{figure}

The structure of single vacancies near the H terminated diamond (100) surface are shown in the Fig.~\ref{fig:Hterm_vacancy_struct}. In these surface vacancy sites there are four $sp^3$ carbon atoms. We try to create a single dangling bond near the surface (the vacancy site 2, 3, 4, 5, 6) by terminating three dangling bonds using hydrogen atoms step by step. Fig.~\ref{fig:vacancy_energy} shows the structure of single hydrogen adsorption on one of the four $sp^3$ carbon atoms. The relative energies are also presented. For the most stable configuration, we then provide more hydrogen atoms to saturate the other dangling bonds. Fig.~\ref{fig:vacancy_saturation} (a) indicates that the energy drops consecutively by 4~eV as a hydrogen atom saturates a dangling bond. Experimental observations indicate that the surface spin can be eliminated via acid and heat treatment. In this hydrogen-vacancy model, the heat treatment will induce hydrogen desorption. Therefore, we calculated the reaction energy barrier of single hydrogen desorption and it yields an energy barrier higher than 2~eV (see Fig.~\ref{fig:vacancy_saturation}). The hydrogen-vacancy model cannot well explain the annealing experiments, thus the models are excluded.

\subsubsection{Locally disordered configurations}

\begin{figure}
\includegraphics[width=8.2cm]{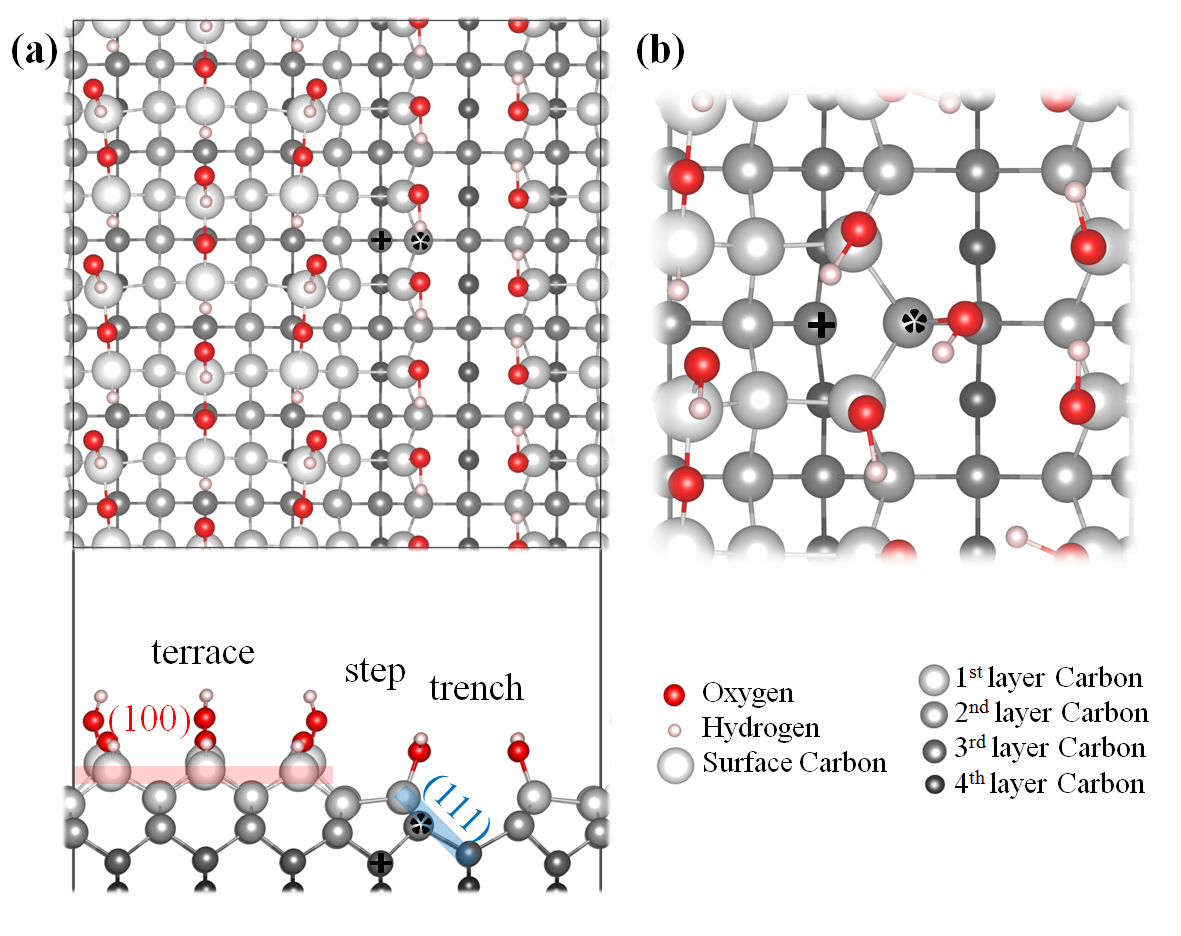}
 \caption{\label{fig:disordered}
   (a) Top and side view of step model of C(100)-6$\times$6 diamond surface. The critical C atoms are labeled by ``+'' and ``*'' before OH adsorption to C(*).
    The (100) and tilted (111) facets are highlighted with red and blue colors, respectively.
   (b) Top view of surface spin model. The DB position occurs for C(+) after adsorption of OH group to C(*) atom.}
\end{figure}

After screening of previous models we conclude that DB defect should be associated with locally disordered configurations. Prior experimental and theoretical work has shown~\cite{Schuelke2013} that the polished oxygenated diamond surface should be disordered to some degree. Our model represents a \emph{locally disordered} configuration on (100) diamond surface. The disorder is modeled by the atomic step on (100) diamond surface that we show in details in Fig.~\ref{fig:disordered}. One carbon at trench site (symbol as ``*'') distorts upwards toward (111) direction once it reacts with atoms or molecules in the environment, and as a consequence, a single $sp^3$ DB defect is formed at the third layer (symbol as ``+''). The floating carbon C(*) at the trench site may be saturated by species in atmosphere, such as an OH radical. X-ray photoelectron spectroscopy and high resolution electron energy loss spectroscopy
 measurements~\cite{gao_water-induced_2008,akhvlediani_interaction_2010} indicate that
 water molecules can dissociate and adsorb on hydrogenated diamond surfaces at room temperature, and these adsorbed water molecules prefer to aggregate on the trench site~\cite{manelli_water_2010}.
 Thus the floating C(*) atom is likely passivated by H or OH. The concentration of C-OH
 on water-rich (100) diamond surface is reduced after annealing to
 600 \,\celsius \ \cite{gao_water-induced_2008,akhvlediani_interaction_2010},
 thus OH desorption is a plausible candidate for surface spin elimination at elevated temperatures.
 Moreover removal of OH from the surface results in no significant other changes at the diamond surface. Therefore, in the absence of obvious surface reconstruction, 
 we assume that the floating C(*) atom is passivated by OH as shown in Fig.~\ref{fig:disordered}(b), 
 and the desorption of OH is the most plausible mechanism for surface spin elimination reaction, 
 which would not significantly alter the surface morphology. 

\begin{figure*}
\includegraphics[width=12cm]{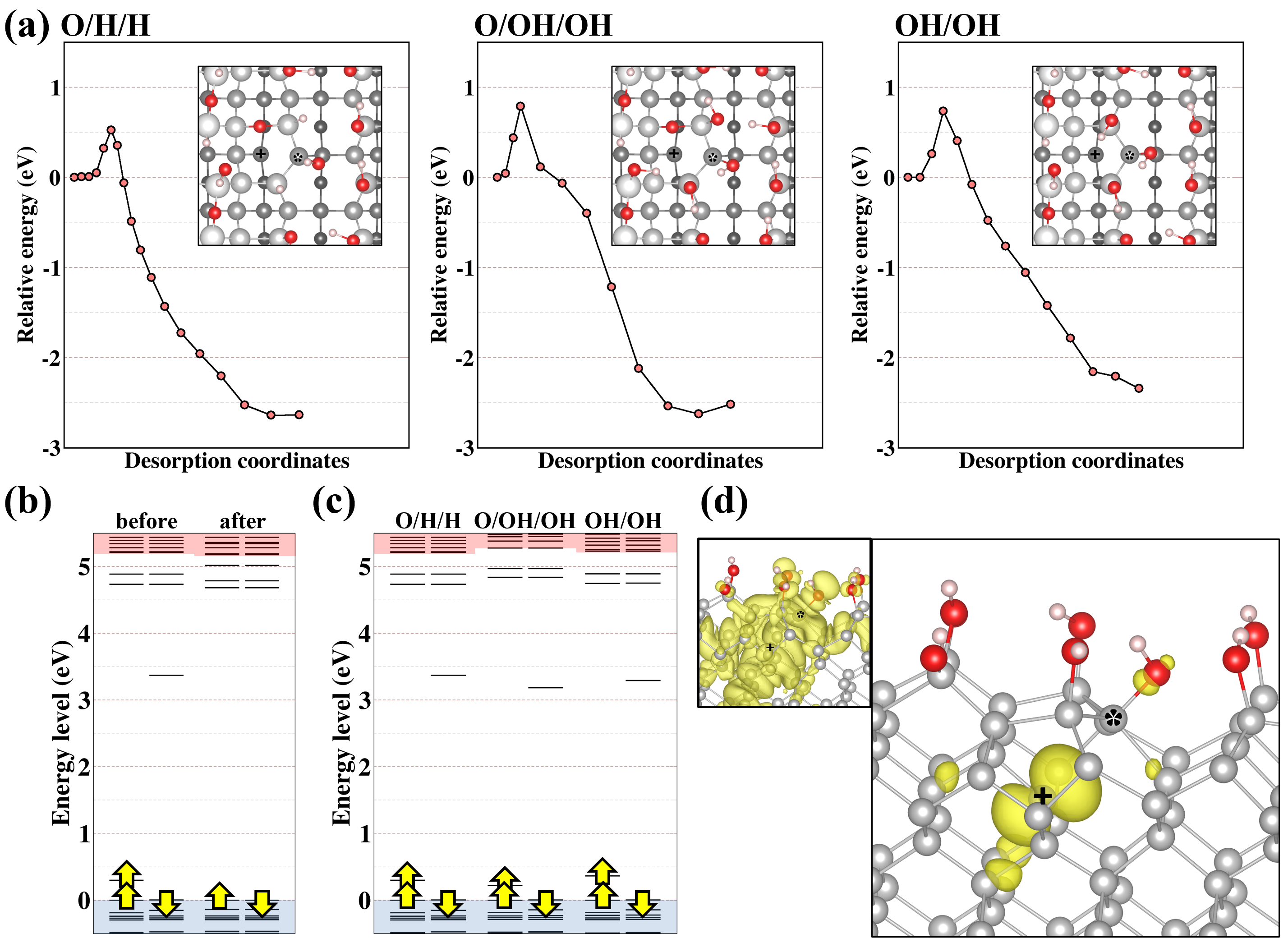}
 \caption{\label{fig:termination} (Color online)
   (a) Top view of three different surface spin defect models. The color settings are the same as those in Fig.~\ref{fig:disordered}.
   (b) The HSE results of the energy levels of O/H/H model before and after OH desorption. 
   (c) The HSE energy levels of three surface spin defect models. The valence and conduction 
    bands are depicted as blue and red regions, respectively. The valence band maximum is aligned to zero. 
   (d) The isosurface of the calculated spin density (isovalues are 5$\times$10$^{2}$ e/Bohr$^3$
    in main figure and 1.3$\times$10$^{4}$ e/Bohr$^3$ in the top-left inset) for surface spin model.
    The spin density is clearly localized on the DB site. The carbon, oxygen, and hydrogen atoms are 
    grey, red, and pink balls, respectively.}
\end{figure*}

We construct a C(100)-6$\times$6 Chadi step model of diamond surface as shown in Fig.~\ref{fig:disordered}, in order to set up a defect forming one DB beneath the top surface. In this model, the surface DBs on the terrace are saturated with H/O/OH groups and the carbon atom on the trench site is saturated by an OH group. The surface spin density in this supercell (area of 15.15$\times$15.15 \AA$^2$) is 4.4$\times$10$^{13}$
 $\mu$B/cm$^2$ which is consistent with experimental estimations of
 $10^{12} \sim 10^{13} \mu$B/cm$^2$~\cite{rosskopf_investigation_2014,myers_probing_2014}, although we note that other defect spins than sp$^3$ DBs may be also present in (100) oxygenated diamond surface~\cite{stacey_evidence_2018}.
 Furthermore, our model implies that even small molecules, e.g., O$_2$ and H$_2$,
 cannot penetrate into the surface step and terminate the DB. The $sp^3$ DB can
 be sterically protected by the top surface of oxygenated diamond under ambient conditions. We emphasize that this prototype model already enables some variations in the vicinity of the topographically protected $sp^3$ DB, e.g., H/OH surface termination around the trench site but it definitely does not cover all the possible variations, e.g., larger atomic steps. To produce statistics about the possible variations would require hundreds of variations which goes beyond the scope of \textit{ab initio} investigations. Nevertheless, a number of variations considered in our study (see Fig.~\ref{fig:termination}) provides an insight about the effect of disorder.

In our proposed model, the surface spin defect is located three atomic layers beneath the (100) diamond. Because the sp$^3$ DB defect is very close the surface and the step edge is a chemically reactive site~\cite{manelli_water_2010}, the local surroundings might alter the physical properties or structural stability of the surface DB defect. Therefore we consider three models, depicting different terminators around the step edge sites. In these models, all C(*) atoms are saturated with OH and the total spin of these systems is $S=1/2$. As shown in Fig.~\ref{fig:termination}(a), (1) O/H/H model, one oxygen is located on the step bridge site and
two trench carbon atoms are saturated by hydrogen atoms; (2) O/OH/OH model, one oxygen is located on the step bridge site and two trench carbon atoms are saturated by hydroxyl radicals; (3) OH/OH model, two trench carbon atoms are saturated by hydroxyl radicals.
Before studying the temperature-dependent OH desorption, it is important to have
a detailed picture of the electronic properties of these surface spin defect models.
The HSE results of O/H/H model before and after OH desorption are shown in Fig.~\ref{fig:termination}(b).
The filled(empty) $sp^3$ DB levels are located at around the position of 0.30(3.3)~eV above valence band maximum ($E_{\mathrm{VBM}}$). After OH desorption, DB states are passivated and only some surface states remain in the bandgap at the position of about 0.7~eV below conduction band minimum (CBM) which are surface C-H and C-OH image states~\cite{kaviani_proper_2014}. The calculated energy levels of these three models are shown in Fig.~\ref{fig:termination}(c). The occupied DB states  of these three models are at $E_{\mathrm{VBM}} + 0.31$, $+0.23$, $+0.38$~eV, respectively, whereas the empty level scatters around $E_{\mathrm{VBM}}+3.3\pm0.1$~eV. As can be seen the adjacent terminators make relatively minor changes to the energy level of $sp^3$ DB states. We note that the absolute values of these levels are subject of supercell size effect and they shift down by about 0.17~eV. The calculated donor and acceptor levels are at $E_{\text{VBM}}+0.42~\mathrm{eV}$ and $E_{\text{VBM}}+3.11~\mathrm{eV}$, respectively.
 
The electron affinity of our oxygenated diamond model is slightly positive at 0.5~eV~\cite{kaviani_proper_2014} which implies that adsorption of electron acceptor molecule
in atmosphere, e.g., water, will cause upward band bending~\cite{ristein_electrochemical_2004}. In realistic oxygenated diamonds this effect may be significant. Recent experiments mapping the band bending in nitrogen implanted (100) oxygenated diamond surface~\cite{broadway_spatial_2018} indeed found a substantial upward band bending under ambient conditions. According to their modeling, the quasi Fermi-level position at the diamond surface is between $E_{\mathrm{VBM}}+1.65~\mathrm{eV}$ and $E_{\mathrm{VBM}}+2.14~\mathrm{eV}$, for implantation depths of nitrogen in the range of $14-70~\mathrm{nm}$, respectively. Taking all the uncertainties in the calculated acceptor levels of $sp^3$ DBs into account one can safely conclude that the $sp^3$ DBs exist in their neutral charge state in realistic oxygenated (100) diamond surface, possessing $S=1/2$ spin configuration. Illumination with green light, i.e., 2.33~eV, typically applied to excite NV center, would not ionize the prototype $sp^3$ DBs in a linear process.

\begin{figure*}
\includegraphics[width=12cm]{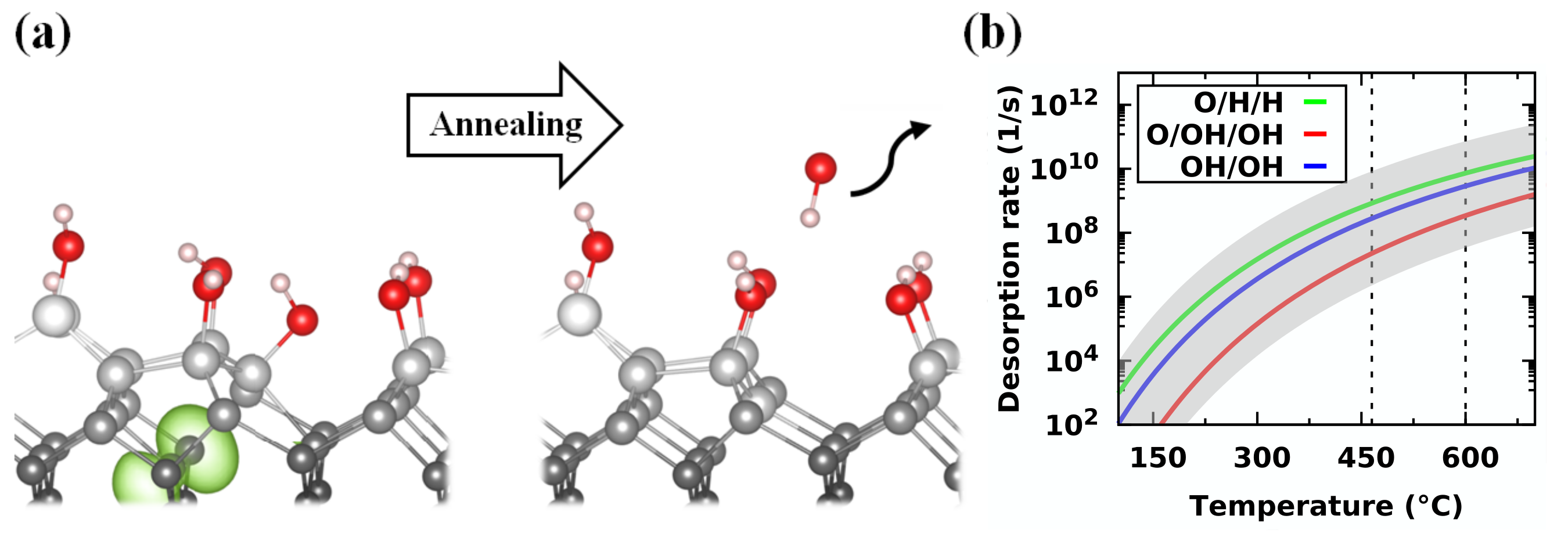}
 \caption{\label{fig:annihilation} (Color online)
   (a) The schematic illustration of surface spin (green lobes) annihilation. 
    After thermal annealing in vacuum, the OH radical desorbs from the surface trench site, causing
    surface reconstruction, and DB is eventually passivated. The color settings are the same as those in Fig.~\ref{fig:disordered}.
   (b) Calculated OH desorption rate of three models as a function of temperature. 
    The annihilation temperatures of 465 and 600~\celsius \ are 
    denoted as dashed lines. The uncertainty in the frequency prefactor is considered by the gray area.}
\end{figure*}

The $sp^3$ DBs can be eliminated by high temperature annealing in O$_2$ atmosphere.
It is computationally prohibitive to model the desorption process in O$_2$ gas, however,
it is possible to carry out annealing in vacuum that can be directly modeled at
\emph{ab initio} level. We propose a spin annihilation mechanism that involves
OH desorption from the spin defect site. To verify our argument, we calculate the OH desorption rate and estimate the number of desorbed OH molecules in a limited area under different temperatures. To calculate the desorption rate, we introduce the Polanyi-Wigner equation~\cite{king_thermal_1975}, 
  \begin{equation}
  \label{eq:desorp}
  R_{\text{des}} = -d\theta / dt = \nu_n \exp ( -E_{\text{des}} / k_{\text{B}} T) \theta_n \text{,}
  \end{equation}
where $E_{\text{des}}$ is the desorption energy barrier, $k_{\text{B}}$ is the Boltzmann constant, $T$ is temperature, $\nu_n$ is the frequency prefactor which is typically $10^{15\pm1}$~1/s due to the larger entropy of the desorbed molecules in vacuum~\cite{alfe_absolute_2006, alfe_ab_2007}.
 $\theta$ is the coverage of OH on DB site, which is equal to one here.
 Because the activation energy barriers of OH desorption would possibly be influenced
 by the local environment, we calculate the energy barriers of the three considered models
 as Fig.~\ref{fig:termination}(a). An illustration of OH thermal desorption from diamond step edge is shown
 in Fig.~\ref{fig:annihilation}(a). The calculated energy barriers in these three models are 0.89, 1.12, and 0.96~eV,
 respectively. Next, we use Eq.~\ref{eq:desorp} to calculate the OH desorption rates
 as a function of temperature as shown in Fig.~\ref{fig:annihilation}(b). The results indicate that
 the number of desorbed OH groups (in one hour) 
 at 600~\celsius \ 
 is at least one order of magnitude more than that at 465~\celsius. Compared with the 
 experimental estimation of surface density ($10^{12} \sim 10^{13}$ $\mu$B/cm$^2$), 
 the desorption energies in the range of 0.89 $\sim$ 1.12~eV result in the removal
 of such amount of surface spins on a surface area of 1 cm$^2$
 at 600~\celsius \ in one hour. At a lower temperature of 465~\celsius \, it requires
 more than three hours to remove all spins in the same area which is consistent with
 the experiment observations~\cite{chu_coherent_2014,lovchinsky_nuclear_2016}.
 We note that an order of magnitude uncertainty in the frequency prefactor translates to a few factors in the annealing times but remain in the right ballpark as shown by the gray area in Fig.~\ref{fig:annihilation}(b).

\subsection{Hyperfine interaction between sp$^3$ dangling bonds and the surface nuclear spins}

So far we have addressed a prototype microscopic structure of diamond surface spins. With a single
electron occupying the near-surface DB the net electron spin is $S=1/2$, 
which means this near-surface DB spin can interact with the present $^{13}$C in diamond or $^1$H isotopes on the surface (nuclear spin $I=1/2$). 
The hyperfine interaction (HFI) between electron spin \textbf{\textit{S}} and a set of nuclear spins \textbf{\textit{I}}$_l$ ($l=1,2,\dots,N$) can be described by the Hamiltonian
 \textbf{\textit{H}}=$\sum$ \textbf{\textit{SA$^{(l)}$I$_l$}} 
 with \textbf{\textit{A$^{(l)}$}} 
 being the hyperfine coupling tensor associated with the $l$th nucleus
 at site \textbf{\textit{R}}$_l$. In atomic units the hyperfine tensor can be written as:
\begin{align}
 \nonumber A_{ij}^{(l)}=&\frac{8\pi}{3} g_e \mu_e g_n \mu_n \rho_n (\textbf{\textit{R}}_l ) \\&+ 
                              g_e \mu_e g_n \mu_n \int d  \textbf{\textit{r}} 
   \frac{3r_i r_j - r^2 \delta_{ij}}{r^5} \rho_s (\textbf{\textbf{\textit{r}}}  ).
\end{align}

 The first term ($a_{iso}$) provides the isotropic hyperfine interaction and
 is referred to as the Fermi-contact term, where $g_{e}(g_{n})$ is the gyromagnetic ratio
 of electron(nucleus), $\mu_e$($\mu_n$) is the Bohr magneton of electron(nucleus), 
 and $\rho_n$($\rho_s$) is the nuclear(electron) spin density. 
 This term is proportional to the magnitude of electron spin density at the nucleus center. 
 The second term provides the anisotropic HFI 
 and is referred to as the dipolar term where \textbf{\textit{r}} is the vector 
 connecting the electron and the nuclear spin. The electron-nucleus Hamiltonian can be 
 put in a simpler form by letting the magnetic field and the crystal \emph{C} axis of 
 symmetry lie in the \textit{xz} plane, where the \textit{C} axis is in the \textit{z} direction. 
 The transformed Hamiltonian is then simplified to 
 $\textbf{\textit{H}} = a\textbf{\textit{S}}_Z \textbf{\textit{I}}_Z
                      + b\textbf{\textit{S}}_Z \textbf{\textit{I}}_X $,
                      \cite{rowan_electron-spin-echo_1965,schweiger_principles_2001}
where the \textit{z} axis is along the applied magnetic field. The quantities 
 $a=A_{ZZ}$ and $b=(A^2_{ZX} + A^2_{ZY})^{1/2}$ 
 describe the secular and pseudo-secular hyperfine couplings. Non-secular terms with 
 $\textbf{\textit{S}}_X$ and $\textbf{\textit{S}}_Y$ are neglected.
 Further, the separation (\textit{r}) between the electron spin and nuclear spin 
 and the angle ($\theta$) that the vector between them makes with applied magnetic field 
 can be extracted from the fitting of HFI parameters ($a$, $b$), where 
 $a = a_{iso} + T(3\cos^2 \theta$$-$$1)$, 
 $b = 3T\sin{\theta} \cos{\theta}$, 
 and $T = g_e \mu_e g_n \mu_n /r^3$ \ \cite{schweiger_principles_2001}. 
 From the above description, it is possible to detect the relative position 
 and angle of a proton around the surface spin.

\renewcommand{\arraystretch}{0.5}
\begin{table}
\caption{\label{tab:hyperfine}Calculated HFI parameters $a$ and $b$ of step models. C(+) and C(*) are DB and floating carbon atoms, respectively, as labeled in Fig.~\ref{fig:disordered}. C labels the other carbon atoms surrounding the carbon dangling bond. $d$ is the distance between C/H and C(+). H$^\prime$ is the hydrogen atom(s) of the adsorbed molecule. H labels the other hydrogen atoms on the surface. Data with either a value or b value larger than 1.0~MHz are presented in this table. The HFI parameters are in MHz unit, and the distances with respect to the DB carbon atom are given in {\AA}.}

\begin{ruledtabular}
\begin{tabular}{cccccccccc}
                    & \multicolumn{3}{c}{O/H/H} & \multicolumn{3}{c}{O/OH/OH} & \multicolumn{3}{c}{OH/OH}  \\
                    &  $d$   &  $a$   &   $b$   &  $d$    &  $a$    &   $b$   &  $d$   &   $a$   &    $b$  \\  \hline
        C(+)        &        & 331.2  &  107.9  &         & 328.5   &  106.3  &        &  327.4  &  106.8  \\
        C(*)        &        &  54.4  &    0.1  &         &  26.0   &    0.4  &        &   26.0  &    0.4  \\  \hline

\multirow{19}{*}{C} &  1.47  &  22.3  &    3.2  &  1.48   &  21.4   &   3.2   &  1.47  &   22.3  &    3.2  \\
                    &  1.48  &  22.9  &    1.4  &  1.49   &  22.0   &   3.0   &  1.47  &   22.5  &    1.7  \\
                    &  1.49  &  22.6  &    3.1  &  1.49   &  21.9   &   1.4   &  1.48  &   22.9  &    2.8  \\
                    &  2.37  &  49.9  &    8.4  &  2.38   &  48.1   &   8.3   &  2.35  &   48.2  &    8.0  \\
                    &  2.37  &  46.4  &    7.8  &  2.38   &  45.1   &   7.9   &  2.37  &   46.5  &    7.8  \\
                    &  2.39  &  14.6  &    1.8  &  2.40   &  10.0   &   2.0   &  2.39  &   10.9  &    2.0  \\
                    &  2.40  &  38.3  &    6.0  &  2.41   &  38.2   &   6.3   &  2.40  &   35.2  &    5.6  \\
                    &  2.48  &   9.6  &    2.1  &  2.48   &   8.3   &   2.0   &  2.42  &   11.6  &    2.1  \\
                    &  2.49  &  11.3  &    1.3  &  2.49   &  14.3   &   1.2   &  2.46  &    7.1  &    1.9  \\
                    &  2.49  &   8.2  &    2.0  &  2.49   &   7.2   &   1.9   &  2.47  &    7.2  &    1.9  \\
                    &  2.51  &  15.6  &    2.8  &  2.50   &  11.6   &   1.9   &  2.52  &   13.4  &    1.3  \\
                    &  2.51  &   9.8  &    2.1  &  2.53   &  11.2   &   1.1   &  2.54  &   13.6  &    1.2  \\
                    &  2.54  &   8.5  &    1.2  &  2.55   &  25.9   &   1.4   &  2.58  &    7.2  &    1.4  \\
                    &  2.78  &   5.2  &    1.7  &  2.80   &  20.4   &   1.6   &  2.60  &    6.9  &    1.3  \\
                    &  2.81  &  11.4  &    0.5  &  2.84   &   4.8   &   1.2   &  2.75  &   38.3  &    3.5  \\
                    &  3.75  &  10.7  &    2.0  &  2.76   &  10.6   &   2.0   &  3.73  &   11.6  &    2.1  \\
                    &  3.76  &  11.3  &    2.0  &  3.76   &  11.8   &   2.0   &  3.74  &   17.4  &    8.1  \\
                    &  3.78  &   6.9  &    1.4  &  3.78   &   6.8   &   1.4   &  3.75  &   10.6  &    1.9  \\
                    &        &        &         &  3.81   &  13.3   &   7.9   &  3.77  &    7.1  &    1.4  \\
\hline
             H'     &  4.22  &   4.6  &    0.2  &  4.15   &   5.5   &   1.7   &  4.15  &    4.4  &    1.4  \\
\hline
\multirow{4}{*}{H}  &  3.55  &  17.6  &    2.2  &  4.23   &   2.4   &   1.2   &  4.06  &    1.9  &    0.8  \\
                    &  3.85  &  14.6  &    2.1  &  4.37   &   1.6   &   0.8   &  4.15  &    2.2  &    1.2  \\
                    &  3.27  &   2.3  &    1.2  &  4.54   &   3.2   &   1.2   &  4.59  &    2.2  &    1.1  \\
                    &  5.90  &   1.0  &    0.8  &  4.70   &   1.4   &   1.2   &  5.75  &    1.1  &    0.9  \\
\end{tabular}
\end{ruledtabular}

\end{table}

Combined with initialization and readout at a proximal NV center, our detailed model
for the hyperfine coupling between the dangling bond and proximate nuclear spins can be used
to probe our model for the dangling bond and to do spectroscopy on surface species. As an example,
we start with the extracted relative positions and angles of two NVs with dark electron spins
in Ref.~\onlinecite{sushkov_magnetic_2014}, labeled as NV$_A$ and NV$_B$.
To compare to the experiment data, we perform calculations of the HFI parameters for our surface spin models with an absorbed OH molecule in different environments, e.g., O/H/H, O/OH/OH, and OH/OH.
The calculated HFI parameters ($a$, $b$) of dangling bond carbon atom and its neighboring atoms are reported in Table~\ref{tab:hyperfine}, the data is sorted by the distance between the DB carbon atom and the neighboring C/H atoms.
By scanning the calculated HFI parameters 
through all C and H atoms, more than ten atoms possess large values (\textgreater \ 10~MHz), 
 but mostly the $b$ values are small (\textless \ 5~MHz). As shown in Fig.~\ref{fig:termination}(d), the spin density is mostly localized
 on the DB site where it reflects a large Fermi-contact value of (329, 105)~MHz on DB $^{13}$C
atom and a small value of about ($4.3-4.6$, $0.2-2.2$)~MHz on surface hydrogen atom of the OH group. 
The latter is in good agreement with an extracted ($a$, $b$) HFI values of a proton for NV$_B$, (4.0, 2.2)~MHz, in Ref.~\onlinecite{sushkov_magnetic_2014}. However, the observed values are associated with the distance between the DB and observed proton at $3.2\pm0.2$~\AA , whereas these hyperfine data are produced with a distance of about 4.1-4.2~\AA\ in the calculation (see Table~\ref{tab:hyperfine}). This clearly demonstrates that oversimplification of the spin Hamiltonian can result in an error of about 30\% in the estimated distance. The atomic step with the trench C(*)-C(+) prototype model indicates that the spin density spreads along the direction of the $sp^3$ DB towards the (111) facet [see Fig.~\ref{fig:termination}(d)]. As a consequence, the largest proton hyperfine parameters are expected to appear for the hydrogen atom which is the part of the chemical group connecting to C(*), i.e., OH group, or their closest neighbors that might be closer to the C(+) atom than the hydrogen atom of the adsorbed OH group  (see Table~\ref{tab:hyperfine}). NV$_A$ with hyperfine couplings of 
around 10~MHz associated with the proton spins in Ref.~\onlinecite{sushkov_magnetic_2014} assumes a hydrogen atom that is within 3~\AA\ within the simple spin dipole model. We find that the order of $10$~MHz hyperfine coupling can be observed for O/H/H model, for which the distance between the protons and the C(+) atom is about 3.6~\AA\ (see  Table~\ref{tab:hyperfine}). Although, the prototype O/H/H model cannot accurately account for the observed hyperfine data but it again demonstrates the need for \textit{ab initio} spin density distribution to accurately determine the hyperfine parameters or measure the distance between the $sp^3$ DB and the protons with relatively short distances. This could be an important issue in the structural analysis of absorbed molecules on the diamond surface by NV quantum sensors. Further investigations 
 using DEER spectroscopy can elucidate alternate possible surface spin configurations, 
 surface proton configurations, and subsurface dark spins. Our results demonstrate that 
 the combination of DEER spectrum and \emph{ab initio} simulations is necessary in the reporter spin protocol for sensing of nuclear spins because the Fermi-contact term can be sizable 
 and no simple approximation is able to estimate its strength. 

\section{Conclusion}
In conclusion, by means of DFT calculations, we propose a simple model of surface spins
on the (100) diamond surface. A (111) facet is essential to create a DB defect beneath diamond
surface and this facet naturally exists at the step of (100) diamond surface. We believe that this prototype model captures the essential components of all the $sp^3$ dangling bond surface defect spins.
We also demonstrated that the OH desorption annihilates the surface spin, and that HFI calculations could provide a great help in the identification of the defect structure in detail.
Our present work indicates large HFI $^{13}$C parameters of about ($a$, $b$)=(337, 106)~MHz
 for the surface dangling bond. A potential future direction is to extend these studies to (111) diamond surface
 where step-free surface can be grown~\cite{tokuda_formation_2012} with proper surface termination.

\begin{figure}[t]
\includegraphics[width=0.4\textwidth]{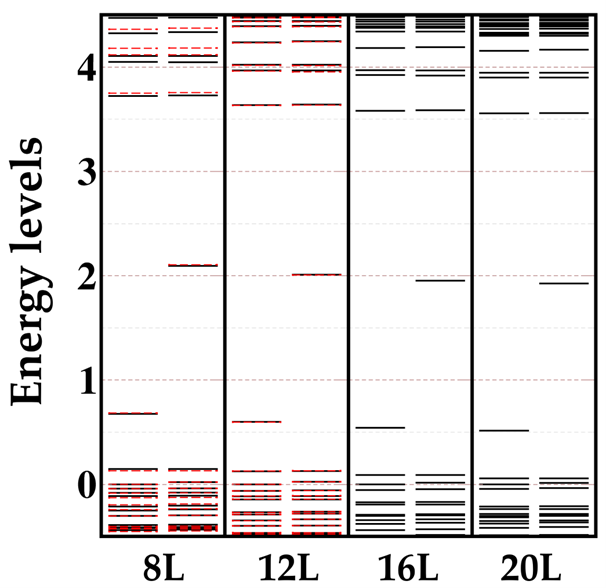}
 \caption{\label{fig:slab} (Color online)
   Energy level plots for the $sp^3$ defect in the function of slab thickness given by the number of layers.}
\end{figure}

\section*{Acknowledgment}
We acknowledge that the results of this research have been achieved
using the DECI resource Eagle HPC based in Poland at Poznan with support from the PRACE aisbl and resources provided by the Hungarian Governmental Information Technology Development Agency (project gallium). A.~G.\ acknowledges the National Research, Development, and Innovation Office of Hungary (NKFIH) grant No.\ KKP129866 of the National Excellence Program of Quantum-coherent materials project and the Quantum Information National Laboratory supported by the Ministry of Culture and Innovation of Hungary (NKFIH grant No.\ 2022-2.1.1-NL-2022-00004) as well as the NKFIH support for the EU QuantERA project MAESTRO and the support from European Commission for the project QuMicro (grant No.\ 101046911). NPdL was supported by the DARPA DRINQS program (grant D18AC00015) and the NSF CAREER program (Grant No. DMR-1752047). J.~C.\ acknowledges the financial support from the Ministry of Science and Technology, Taiwan (MOST 109-2112-M-018-008-MY3).

\section*{Appendix}
\subsection*{Slab model and vacuum size convergence tests}

We calculate the Kohn-Sham energy levels for diamond (100) $sp^3$ DB surfaces with different size of layer thickness and vacuum region as shown in Fig.~\ref{fig:slab}. We carried out this test with the computationally affordable PBE DFT functional which does not reproduce the experimental band gap of diamond. Therefore, the shift in the Kohn-Sham energy levels as a function of the system size can be read out from the plot and not the absolute position with respect to the valence band edge or vacuum level. The black bars are energy levels of 8L, 12L, 16L, and 20L thickness that the cell are fixed thus the corresponding vacuum sizes are 27.3 {\AA}, 23.8 {\AA}, 20.2 {\AA}, and 16.6 {\AA}. The results indicate that the maximum energy deviation is 0.17~eV with shifting down the DB Kohn-Sham level. We also evaluate the energy deviation for different vacuum sizes as indicated by the red dashed bars. The vacuum sizes are changed from 27.3 {\AA} to 12 {\AA} for 8L case and from 23.8 {\AA} to 27 {\AA} for 12L case, the energy deviation are both less than 20~meV.

\bibliography{reference}

\end{document}